# Observation of Little-Parks oscillations at temperatures much lower than the critical temperature using a GHz resonator

A. Belkin[*], M. Brenner[*], T. Aref, J. Ku, and A. Bezryadin

*Department of Physics*
*University of Illinois at Urbana-Champaign*
*Urbana, IL 61801*

**Abstract:**

*It is demonstrated that a thin-film Fabry-Perot superconducting resonator can be used to reveal the Little-Parks (LP) effect even at temperatures much lower than the critical temperature. A pair of parallel nanowires is incorporated into the resonator at the point of a supercurrent antinode. As magnetic field is ramped, the Meissner current develops, changing the kinetic inductance of the wires and, correspondingly, the resonance frequency of the resonator, which can be detected. The LP oscillation are revealed as a periodic set of distorted parabolas observed in the transmission of the resonator and corresponding to the states of the wire loop having different vorticities. We also report a direct observation of single and double phase slip events and their statistical analysis.*

The phenomenon of magnetic flux quantization in doubly connected superconductors was observed by Deaver and Fairbank[1] and Doll and Nabauer[2]. In these first works, authors investigated the magnetic properties of thick (wall thickness is much larger than magnetic penetration depth) superconducting tubes. Besides the proof of the flux quantization phenomenon both experiments showed that the magnetic flux quantum period was not *hc/e*, but rather *hc/2e*. In mesoscopic samples the flux quantization is irrelevant since their dimensions are typically smaller than the magnetic field penetration depth. In this case the principle of fluxoid quantization, which follows from the requirement that the complex wavefunction of the condensate is single-valued, holds. The experimental demonstration of the fluxoid quantization was first reported by Little and Parks[3] (LP) who demonstrated that the critical temperature of a thin-walled superconducting cylinder is a periodic function of the magnetic flux contained in the cylinder. The phase diagram of the thin hollow (wall thickness is much smaller than magnetic penetration depth) cylindrical superconductor, obtained by LP from multiple resistance vs. temperature (R-T) measurements revealed the presence of clearly defined series of parabolic variations of the critical temperature with magnetic field[4,5]. More recently, Vakaryuk[6] predicted that at low temperatures the magnetic moment of a superconducting loop should oscillate with the applied field either with the LP period or with a doubled period. This theory can be considered an extension of the LP theory to mesoscopic samples and low temperatures.

So far measurements of the LP effect were mostly concentrated near the critical temperature where the resistance is still high, due to thermal fluctuations for example. At low temperatures the observation of the LP oscillations is challenging since the resistance is immeasurably low. Here we measure the kinetic inductance of a pair of parallel wires, which, together with superconducting electrodes, form a closed loop. These wires and electrodes constitute a superconducting coplanar waveguide resonator (Fig. 1 (a)). The kinetic inductance variations reveal themselves as changes of the resonance frequency, which, in their turn, produce changes of the transmission of the resonator, measured at a fixed frequency, which equals the resonance frequency at zero field. We show that the transmission coefficient ($S_{21} = 10 \log(P_{out} / P_{in})$, where $P_{out}$ and $P_{in}$ are output and input powers, correspondingly) of the resonator has multiple branches as a function of the external magnetic field. The transition from one branch to the neighbor branch corresponds to a single Little's phase slips[7], taking place in one of the wires. The periodic structure of the $S_{21}(H)$ dependence has the same origin as the Little-Parks critical temperature $T_c(H)$ periodic oscillations, occurring with changing of the applied magnetic field $H$. These oscillations are due to the oscillation of the supercurrent magnitude and, correspondingly the free energy of the superconducting condensate with magnetic flux, caused by a periodic entrance of vortices into the superconducting loop.

---

[*] These authors contributed to the work equally.

Our device consists of a pair of parallel superconducting nanowires incorporated in the center of a superconducting coplanar waveguide resonator (Fig.1 (a)). The nanowires (Fig.1 (b), (c)) with thickness ~ 25 nm and ~ 100 nm length were produced by molecular templating technique[8,9]: a carbon nanotube was deposited over a 100 nm wide trench on a Si substrate and sputter-coated with a superconducting alloy of MoGe. Resonators were patterned by means of optical lithography, followed by wet chemical etching in $H_2O_2$ (Fig.1 (a)). The width of the center conductor is 10 μm and the gap between the center conductor and ground plane is 6 μm, the gap between the center conductor and the input (or the output) electrode is 3 μm with corresponding capacitances of about 45 fF. The length of the center conductor between the input and the output "mirrors" is 10 mm, which corresponds to a nominal resonant frequency ~ 10 GHz (taking into account the dielectric constant of the underlying substrate), which is estimated for the case when the kinetic inductances of the resonator itself and the nanowires are both negligible. The actual resonance frequency is lower than this simple estimation, due to the kinetic inductance of the thin MoGe film forming the resonator as well as the nanowires. Two samples are measured with different nanowire spacing. In sample A, the distance between nanowires is 6.63 μm, and in sample B the distance between nanowires is 4.26 μm. The resonators are over-coupled and their quality factors are about 500. Each measurement is performed in a $He^3$ cryostat equipped with two semi-rigid coaxial cables with thermalized attenuators and a low temperature GHz amplifier LNF-LNC4_8A, from Low Noise Factory. The transmission measurements are performed using a network analyzer Agilent PNA5230A.

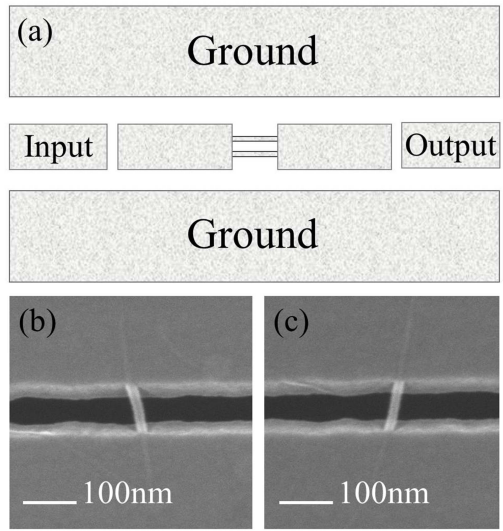

Figure 1 (*a*) The schematics of the studied resonator-nanowire-loop device, with two nanowires in the center. The grey texture corresponds to the $Mo_{76}Ge_{24}$ film. White regions are the regions where the MoGe film was removed. The center conductor is capacitively coupled to the input and output electrodes. (*b*), (*c*) SEM images of nanowires in the center of resonator, sample A. "Bright spots" are visible at the ends of the wires, indicating that the nanowires are straight and they do not "dive" into the trench.

Our measurements, shown in Fig. 2 (a) for sample A and in Fig. 3 for sample B, reveal that $S_{21}(H)$ is a periodic multi-valued function. Each branch of $S_{21}(H)$ corresponds to a state with a particular number of phase vortices (or vorticity) trapped in the nanowire loop (Fig.1 (a)).

A single branch of the function $S_{21}(H)$ (for the sample A) obtained at 0.36 K is extracted and plotted in Fig. 2 (b). As one can see, the transmission function is not truly parabolic but rather it has a flat top. Therefore we will call it a "distorted-parabola". This profile is preserved in a wide temperature interval. For example, the distorted-parabola transmission function is observed on sample B at 0.3 K and 1.8 K, as shown in Fig. 3.

To examine the periodicity of the transmission coefficient in the external magnetic field we compare the mean distance in magnetic field between the intersections of branches with the following vorticities: $(n,n+1)$ (not shown), $(n,n+2)$ (shown as circles in Fig. 2 (a)) and $(n,n+3)$ (shown as triangles Fig. 2 (a)). This analysis results in the following periodicities: $\langle \Delta H_{n,n+1} \rangle = 0.420$ Oe, $\langle \Delta H_{n,n+2} \rangle = 0.423$ Oe, $\langle \Delta H_{n,n+3} \rangle = 0.424$ Oe, and the mean value of $\langle \Delta H(0.36 \text{ K}) \rangle = 0.422$ Oe. A similar analysis for this sample at 1.80 K (not shown) results in the following values for the period: $\langle \Delta H_{n,n+1} \rangle = 0.427$ Oe, $\langle \Delta H_{n,n+2} \rangle = 0.427$ Oe and $\langle \Delta H_{n,n+3} \rangle = 0.424$ Oe. This yields the mean value of $\langle \Delta H(1.80 \text{ K}) \rangle = 0.426$ Oe. Consequently, the difference between periods of the transmission coefficient function at 0.36 and 1.80 K is less than 1%, which is within the uncertainty of the measurement. Thus we conclude that the period is independent of the temperature in a wide temperature interval. This result agrees with the previous conclusion about independence of the magnetic-field-induced LP oscillations on temperature in systems with two parallel wires connected to wider superconducting electrodes[10]. The value of the period in a magnetic field allows one to establish the relationship between the applied magnetic field and the phase change of the superconducting order parameter (condensate wave function) of the studied double-wire system. Indeed, the period of Little-Parks oscillations corresponds to the phase change of $2\pi$ on the closed loop: $\Delta H \cdot \beta = 2\pi$ from where we find the field-phase geometrical parameter $\beta \approx 14.8$ Oe$^{-1}$. We have also tested that the observed multivalued response function is independent on the input power. The powers tested were -60, -70, and -80 dBm, referring to the output of the network analyzer. This is four, five and six orders of magnitude lower than the critical power at which the current amplitude in the nanowires reaches the critical current[11]. The observed independence of the measured $S_{21}(H)$ on the input power proves that the measurement current (i.e. the induced oscillating current in the resonator) is negligible compared to the magnetic-field-induced Meissner current in the nanowire loop.

If the magnetic field is exactly perpendicular to the sample's surface then the period is given by[10] $\Delta B = (\pi^2/8G) \cdot (\Phi_0 /al)$, where $a$ is the distance between the wires, $l$ is the width of the electrodes to which the wires are connected and $G = 0.916$ is Catalan number[12] (the formula is exactly correct only if $l \gg a$). In the experiment the angle between the magnetic field and the sample's surface was $\theta \approx 35\text{-}40°$. The corresponding expression for the period is $\Delta B = (\pi^2/8G) \cdot (\Phi_0 /al \sin\theta)$. With the wire separation $a = 6.63$ μm and the electrode width $l = 10$ μm, we get $\Delta B = 0.59$ to $0.66$ Oe. This is close to the experimentally measured vale of $\Delta B = 0.43$ Oe. The difference is due to the fact that the condition $l \gg a$ is not fulfilled.

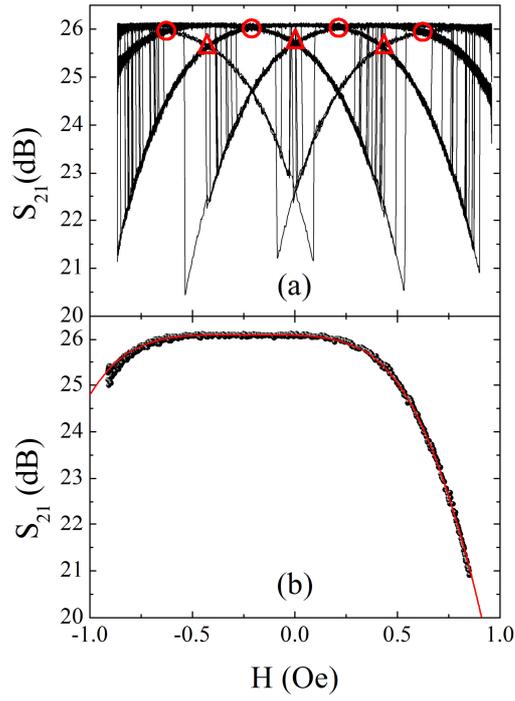

*Figure 2. (a) The transmission coefficient of sample A as a function of the external magnetic field, measured at 360 mK. Different branches correspond to different numbers of phase vortices trapped in the loop formed by the nanowires and the electrodes (see Fig. 1 (a)) The data points are shown by dots connected by straight lines, thus the vertical lines represent abrupt jumps from one parabola to another. Circles and triangles represent the intersections of branches with the vorticities (n,n+2) and (n,n+3), correspondingly. (b) A single branch of the experimental $S_{21}(H)$ dependence (black dots). Red line shows the theoretical fit. An excellent agreement is observed.*

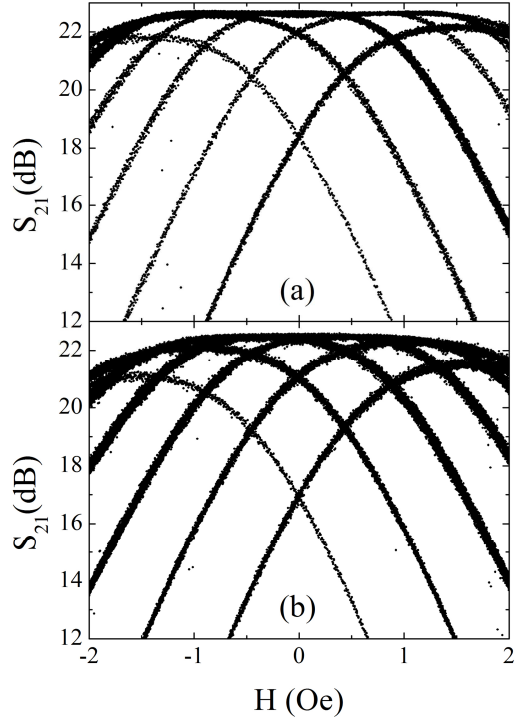

*Figure 3. The transmission coefficient for sample B as a function of the external magnetic field at two temperatures (a) 0.3 K and (b) 1.8 K. The output power of the network analyzer is -60 dBm.*

The transmission coefficient of a coplanar waveguide resonator depends on the resonant frequency $f_0$, the probe frequency $f$, and the resonator's quality factor $Q$ and is described by Lorentzian:

$$S_{21} = 10\log\left[\frac{A_0}{(f_0/Q)^2 + 4(f-f_0)^2}\right]$$

where $A_0$ is some constant. In a magnetic field the resonance frequency of a superconducting resonator changes (due to the changes of the kinetic inductance of the wires) and the field-dependent transmission coefficient reads:

$$S_{21}(H) = S_{21}(0) - 10\log\left[1 + 4\left(\frac{f_0' - f_0'}{f_0}Q\right)^2\right]$$

where $f_0'$ is the field-dependent resonance frequency. The latter expression for $S_{21}(H)$ is derived under the assumption that the probe signal is not changed and fixed at $f_0$, which is the resonance frequency in zero magnetic field. It is also assumed that the applied field is so weak that the quality factor of a resonator does not depend on the field. The above expression is used to generate the theoretical fit in Fig. 2 (b) (red curve). The total inductance ($L$) of the sample can be estimated as a sum of the inductance of a resonator itself ($L_{res}$) and the kinetic inductance of the nanowires ($L_{nw}$), $L=L_{res}+L_{nw}$. The kinetic inductance of a nanowire is due to the inertia of the moving condensate (we neglect the magnetic inductance of the nanowire due to its small dimensions and short length). The kinetic inductance of the wire depends on its current-phase relationship (CPR) and is given by the following expression

$$L_{nw} = \frac{\hbar}{2e}\left(\frac{dI}{d\phi}\right)^{-1}$$

Here $e$ is the electronic charge, $\hbar$ is the reduced Planck's constant, $\phi$ is the phase difference between the ends of the wire, and $I$ is the supercurrent in the wire. We have performed numerical simulations using the Likharev[13] CPR expression for a long wire ($l \gg \xi$, where $l$ is the length of the nanowire and $\xi$ is the coherence length):

$$I = \frac{3\sqrt{3}}{2} I_C \left( \frac{\phi}{(l/\xi)} - \left(\frac{\phi}{(l/\xi)}\right)^3 \right)$$

Here $I_C$ is the critical current of the wire. From the two previous formulas the inductance of the nanowire is:

$$L_{nw}(H) = \frac{\hbar}{3\sqrt{3}e} \frac{1}{I_{C1}+I_{C2}} \frac{1}{\frac{1}{(l/\xi)} - 3\frac{(\beta H/2)^2}{(l/\xi)^3}}$$

Using the expression for the resonant frequency $f_0 = (4\pi^2 LC)^{-1/2}$, we obtain the shift of the resonance peak in the magnetic field:

$$f_0 - f_0' = \frac{1}{2\pi C_{res}^{1/2}} \left( \frac{1}{\sqrt{L_{res}+L_{nw}(0)}} - \frac{1}{\sqrt{L_{res}+L_{nw}(H)}} \right)$$

where $C_{res}$ is the capacitance of the resonator with nanowires. In this formula we have assumed that $\phi = \beta H/2$ where the factor 2 originates from the fact that the total phase difference generated on the wire loop by the applied magnetic field is shared, presumably equally, between the two wires. Using these expressions we are able to fit experimental data with the fitting parameters $I_{C1} = I_{C2} = 33$ µA, $l/\xi = 20$, $L = 6$ nH and $\beta = 12$. The result is illustrated in Fig. 2 (b) (red curve).

The experimental approach outlined here allows one to tell exactly how many vortices entered the loop. The vertical lines in Fig. 2 (a) represent the events when the vorticity changes. The vorticity changes almost instantaneously. The vertical lines in Fig. 2 (a) appear because in this figure dots representing data are connected by straight lines. Thus the lines show from which parabola to which other parabola the jump takes place. If the line connects two neighboring parabolas then one vortex has entered the loop. If the line connects the next to the near neighbor parabola then two vortices have entered, meaning a double phase slip has occurred, etc.

It was theoretically predicted[14] that transitions between states with different vorticities have unequal probabilities. Under certain conditions the entrance of two vortices at once into a superconducting loop are expected to be more probable. Analysis of our data reveals that in the temperature range $T < 1.5$ K the rate of transition with double phase slip events is significantly higher than rates with single or triple phase slip events (see Fig. 4). As we increase the temperature above 1.5 K the picture changes, namely the frequency of the $n \rightarrow n+2$ jumps decreases while the frequency of the $n \rightarrow n+1$ jumps increases. At approximately 2 K these two rates become equal and $n \rightarrow n+3$ transitions become very rare. Further rise of temperature leaves only $n \rightarrow n+1$ transition possible. This phenomenon is not yet fully understood.

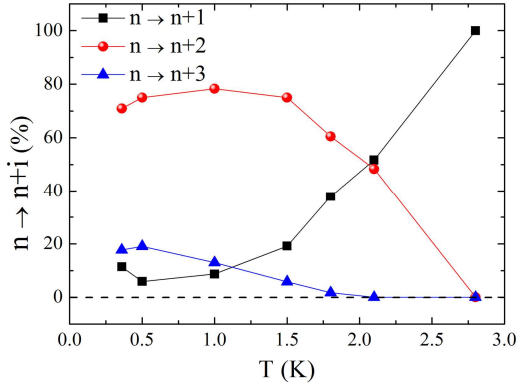

*Figure 4. Relative frequencies for transitions of the type n→n+1, n→n+2, and n→n+3, calculated for 85 transitions. Here n is the vorticity, i.e. the number of the phase vortices present in the loop.*

The jumps between parabolas have a one-to-one correspondence with phase slips. Based on our measurement we are able to provide an upper bound to the rate of quantum phase slips (QPS) in nanowires[15]. The conclusion is that in the linear regime the rate of quantum phase slips is extremely low and could be zero. Let us explain this in more detail. The flat top of the parabola (Fig. 2 (b)) corresponds to a "linear regime", i.e. such regime when the Meissner current in the loop is weak enough so the kinetic inductance of the wires is not changing significantly. For the single parabola shown in Fig. 2 (b) the linear regime roughly extends from -0.5 Oe to 0.25 Oe. Considering all the data we have obtained on both samples, we conclude that no phase slips ever happen in the linear regime. Experimentally we find that the jumps from one parabola to another can only happen if the field is such that the circulating current is strong enough to significantly change the wire kinetic inductance (i.e. in the nonlinear regime). The time it takes the field to sweep near the flat top of the distorted parabolas is ~ 50 s. Since no phase slips are ever observed near the top of the parabola, the rate of QPS is lower than 0.02 s$^{-1}$. Taking into account the fact that each sample was measured at least 100 times, one concludes that the QPS rate is less than $2 \cdot 10^{-4}$ s$^{-1}$. This should be compared to the rates predicted by Golubev and Zaikin (GZ)[15], in their Table 1. In our sample A wires have the width ~ 22 nm, the thickness ~25nm and the length of the order of 100 nm, consequently their resistance[9] ~ 0.45 kOhm. Thus the ratio resistance/length is ~ 0.0045 kOhm/nm. From the equations 47 and 50 of the GZ paper, one can roughly estimate the rate of phase slips as $10^{-115}$ s$^{-1}$ for A = 1, where A is an unknown coefficient introduced by Golubev and Zaikin (we imply that the attempt frequency, *B*, is equal to the resonant frequency of the resonator, i.e. ~ 5 GHz). This estimate is consistent with our result. A qualitative conclusion is that in a MoGe wire with the diameter as small as ~22 nm the QPS do not occur and the wire stays coherent, on a scale of ~ 1 hour at least, unless a strong current is applied. The robustness of the condensate was even stronger in sample B having the wires with similar width ~ 26nm and ~ 25 nm thick. In this case to initiate the jumps between parabolas and to obtain the results shown on Fig. 3, we had to apply microwave pulses every 100 seconds (the measurements were done after the application of each pulse but not during the pulse).

Generally speaking, we classify the observed periodic set of the distorted parabolas as a manifestation of the Little-Parks effect because it is a reflection of the underlying fact that the thermodynamic potential of our loop formed by wires and electrodes is a periodic set of parabola-like curves, originating from the fundamental principle – single-valuedness of the condensate wave function and the corresponding quantization of the phase increase along the closed path by 2π·n (*n* is an integer). The same periodic set of parabolas describing the free energy thermodynamic potential is the underlying physical mechanism that leads to the classic Little-Parks oscillation of the critical temperature, observed on empty cylinders[3].

In conclusion, we have shown that a coplanar waveguide resonator can be used to study the Little-Parks periodicity at low temperatures and at low bias. The LP periodicity can be seen in the dependence of the transmission function on the magnetic field as a set of periodically shifted distorted parabolas. Each parabola represents a certain vorticity of the system. Jumps between parabolas allow us to distinguish between single, double and triple phase slip events. The jumps in which the vorticity changes by two (i.e. two vortices enter the closed nanowire loop in one jump) are more frequent at low temperatures.

The work was supported by the DOE grant DE-FG02-07ER46453.